\documentclass[aps,prl,twocolumn,superscriptaddress,showpacs,floatfix]{revtex4}
\usepackage{graphicx,epsfig,amsmath}

\newcommand{\bohrm}{\mbox{$\mu_{B}$}}
\newcommand{\kboltz}{\mbox{$k_{B}$}}

\newcommand{\etal}{\textit{et al.}}

\newcommand{\Tone}{\mbox{$\text{T}_{\text{1}}$}}

\newcommand{\ZeemanE}{\mbox{$|g|\bohrm B$}}

\newcommand{\startsubfig}[2]{Figure~\ref{fig:#1}(#2)}
\newcommand{\subfig}[2]{Fig.~\ref{fig:#1}(#2)}

\newcommand{\AsubB}[2]{\mbox{$\text{#1}_{\text{#2}}$}}

\begin{document}


\title{Measurements of the spin relaxation rate at low magnetic fields in a quantum dot }

\author{S. Amasha}
	\email{samasha@mit.edu}
	\affiliation{Department of Physics, Massachusetts Institute of Technology, Cambridge, Massachusetts 02139}

\author{K. MacLean}
	\affiliation{Department of Physics, Massachusetts Institute of Technology, Cambridge, Massachusetts 02139}

\author{Iuliana Radu}
	\affiliation{Department of Physics, Massachusetts Institute of Technology, Cambridge, Massachusetts 02139}

\author{D. M. Zumb\"{u}hl }
	\affiliation{Department of Physics, Massachusetts Institute of Technology, Cambridge, Massachusetts 02139}
	\affiliation{Department of Physics and Astronomy, University of Basel, Klingelbergstrasse 82, CH-4056 Basel, Switzerland}

\author{M. A. Kastner} 
	\affiliation{Department of Physics, Massachusetts Institute of Technology, Cambridge, Massachusetts 02139}
	
\author{M. P. Hanson} 
	\affiliation{Materials Department, University of California, Santa Barbara 93106-5050}	

\author{A. C. Gossard} 
	\affiliation{Materials Department, University of California, Santa Barbara 93106-5050}


\begin{abstract}

	We measure the relaxation rate $W \equiv \Tone^{-1}$~of a single electron spin in a quantum dot at magnetic fields from $7$ T down to $1.75$ T, much lower than previously measured. At $1.75$ T we find that \Tone~ is $170$ ms. We find good agreement between our measurements and theoretical predictions of the relaxation rate caused by the spin-orbit interaction, demonstrating that spin-orbit coupling can account for spin relaxation in quantum dots. 

\end{abstract}

\pacs{73.63.Kv, 03.67.Lx, 76.30.-v}

\maketitle


	To implement proposals \cite{Loss1998:QDotQuanComp} for quantum computation based on manipulating electron spins \cite{Petta2005:CoherentManipulation} in quantum dots \cite{Kouwenhoven1997:NatoReview}, the spin of the electron must remain coherent for a sufficiently long period of time. One decoherence mechanism that can affect this time is spin-orbit coupling \cite{Golovach2004:PhononInducedDecay}. In a magnetic field $B$ the spin states of a single electron in a dot are split by the Zeeman energy $\Delta= \ZeemanE$. The energy relaxation time \Tone~ is the average time necessary for an electron in the excited spin state to relax to the ground spin state. In single quantum dots, it has been predicted \cite{Khaetskii2001:ZeemanSF,Golovach2004:PhononInducedDecay,Falko2005:Anisotropy} that spin relaxation is caused by the spin-orbit interaction and that \Tone~increases with decreasing magnetic field. Pulsed gate transport measurements \cite{Fujisawa2001:EnergyRelaxation, Hanson2003:SpinRelaxation} have put lower bounds on \Tone, while Elzerman \etal~\cite{Elzerman2004:SingleShotReadOut} have utilized an energy-selective spin readout technique to measure \Tone~ for one electron in a single dot at large magnetic fields and found $\Tone = 0.85$ ms at $B= 8$ T. Hanson \etal~\cite{Hanson2005:SpinDepTunnelRates} have measured the singlet-triplet relaxation time at smaller fields for two electrons.
	
	In this Letter, we present measurements of the relaxation rate $W \equiv \Tone^{-1}$ of one electron in a single dot at magnetic fields from $7$ T down to $1.75$ T, much lower than previously measured. These measurements are possible because of the good stability of the heterostructure we used combined with an active feedback system that compensates for residual drift and switches of the dot energy levels, allowing us to measure down to fields where $\Delta$ is comparable to our electron temperature. We find relaxation times as long as $170$ ms at $1.75$ T. We compare our measurements of $W$ vs $B$ to theoretical predictions by Golovach \etal~\cite{Golovach2004:PhononInducedDecay} of the relaxation rate caused by spin-orbit coupling and find excellent agreement between theory and experiment. This demonstrates that spin-orbit coupling can account for the relaxation of the spin of a single electron in a quantum dot.

\begin{figure}[!]
\setlength{\unitlength}{1cm}
\begin{center}
\begin{picture}(8,10)(0,0)

\put(0,0){\includegraphics[width=8.0cm, keepaspectratio=true]{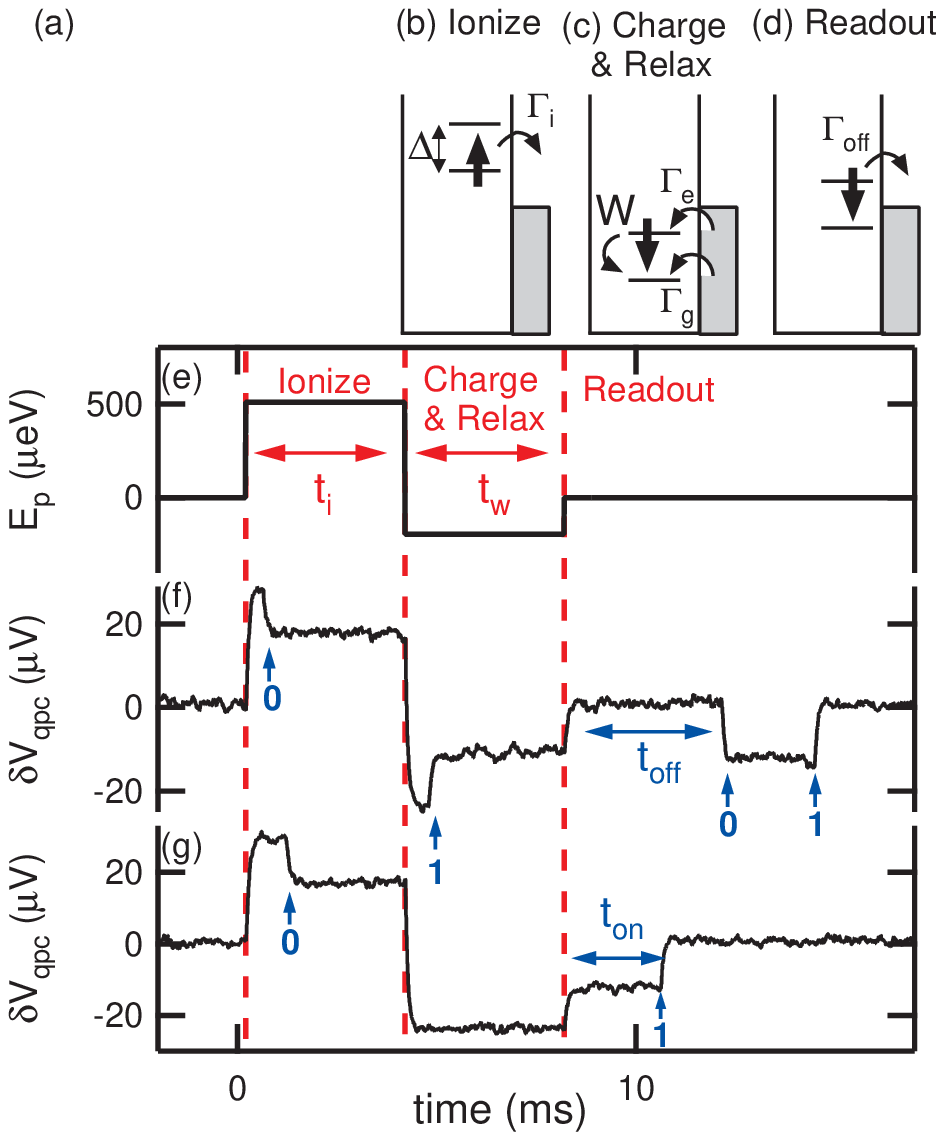}}
\put(0,6.9){\includegraphics[width=3.3cm,keepaspectratio=true]{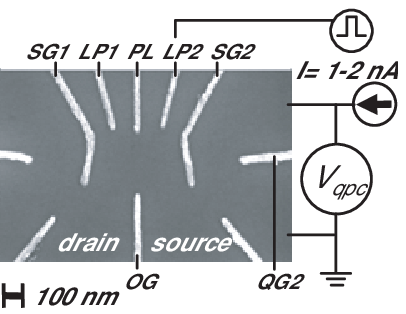}}

\end{picture}
\end{center}

	\caption{(a) Electron micrograph of the gate geometry. Negative voltages are applied to the labeled gates to form the quantum dot and the QPC charge sensor; unlabeled gates are grounded. Pulses are applied to gate LP2. The drain and source electrodes are labeled and are grounded. (b)-(d) diagrams showing the configuration of energy levels for the three steps in the pulse sequence shown in (e). (f) and (g) are examples of data taken at $B= 2.5$ T and $t_{w}= 4$ ms. The direct capacitive coupling between LP2 and the QPC causes the QPC to respond to the pulse sequence; electron tunneling events are evident on top of this response. The 0's denote when an electron tunnels off the dot, while 1's denote when an electron tunnels on. 
}
	\label{fig:pulseseq}
\end{figure}

	The dot used in this work is fabricated from an AlGaAs/GaAs heterostructure. The two-dimensional electron gas (2DEG) formed at the AlGaAs/GaAs interface $110$ nm below the surface has an electron density of $2.2\times10^{11}$ $\text{cm}^{-2}$ and a mobility of $6.4\times10^{5}$ $\text{cm}^{2}$/Vs \cite{Granger2005:TwoStageKondo}. The gate geometry is shown in \subfig{pulseseq}{a} and is based on that of Ciorga \etal~\cite{Ciorga2000:AdditionSpectrum}. We choose gate voltages so that we form a single dot containing one electron. For this work we have tuned the barrier formed by the gates OG and SG1 to have a tunnel rate much lower than the rate through the barrier defined by OG and SG2. We measure the dot in a dilution refrigerator with an electron temperature of about $120$ mK. To minimize orbital effects we align the 2DEG parallel to the magnetic field to within a few degrees.
		
	We use the quantum point contact (QPC) formed by SG2 and QG2 as a sensitive electrometer or charge sensor \cite{Field1993:NoninvasiveProbe} for the dot. The detection circuit is illustrated in \subfig{pulseseq}{a} and more details are in Refs. \cite{Amasha2006:TowardManip,Amasha2006:TunnelingSummary,MacLean2006:WKB}. If an electron tunnels onto or off the dot, it changes the electrochemical potential of the electrons in the QPC, which in turn causes a change in resistance $\delta R$. We observe $\delta R$ by sourcing a $1-2$ nA current across the QPC and measuring the change in voltage $\delta \AsubB{V}{QPC}$. By making the tunneling rate through the OG-SG2 barrier less than the bandwidth of our circuit, we observe the electron tunneling in real time \cite{Elzerman2004:SingleShotReadOut,Lu2003:RealTimeDet,Schleser2004:TimeResolvedDet}. Our typical signal size of $10 \mu$V is approximately $5\%$ of the total voltage across the QPC; this good sensitivity may come from making the gate SG2 between the dot and the QPC narrow, which increases the coupling between the dot and the QPC \cite{Zhang2004:EngineerQPC}. The small QPC current does not heat the electrons, which is important because it is our electron temperature which sets the lower limit on the fields we can measure.       

	To measure $W$ at a given magnetic field, we apply a three step pulse sequence \cite{Elzerman2004:SingleShotReadOut} $V_{p}$ on top of the dc voltage on gate LP2: $\AsubB{V}{LP2}= \AsubB{V}{dc} + \AsubB{V}{p}$. This sequence is illustrated in \subfig{pulseseq}{e}, where we have converted the gate voltage pulse \AsubB{V}{p}~into the equivalent electrochemical potential energy change of the dot: $\AsubB{E}{p} = -e\alpha_{LP2} \AsubB{V}{p}$, where $\alpha_{LP2}\approx 0.065$ is the capacitance ratio for gate LP2 extracted from transport measurements \cite{Amasha2006:TunnelingSummary}. The first step is to apply a negative \AsubB{V}{p}~ to bring both spin states above the Fermi energy of the lead, as shown in \subfig{pulseseq}{b}. We hold the dot in this configuration for a fixed time $t_{i}$, during which time the electron can tunnel off the dot. After time $t_{i}$ the dot is in one of three possible states: there is a probability $P_{i}$ that the dot is ionized given by the ionization efficiency $\epsilon_{i}= 1-e^{-\Gamma_{i} t_{i}}$, where $\Gamma_{i}$ is the tunnel-off rate. In this work, $\epsilon_{i} \approx 0.95$. The probability of being in the ground state is $P_{g} \approx 1-\epsilon_{i}$. Finally, the probability of being in the excited state is thermally suppressed and is $P_{e} = e^{-\Delta/\kboltz T} P_{g}$, which is negligible. 
	
	After ionizing the dot, we apply a positive \AsubB{V}{p}~ and bring both states below the Fermi energy of the lead as shown in \subfig{pulseseq}{c}. We hold the dot in this configuration for a time $t_{w}$, which we vary. During this time, electrons tunnel into either the ground or excited states of the dot with rates $\Gamma_{g}$ and $\Gamma_{e}$, respectively. We expect that $\Gamma_{e} / \Gamma_{t} = 0.5$ where $\Gamma_{t}= \Gamma_{e}+ \Gamma_{g}$, but we do not assume this a-priori, since we extract $\Gamma_{e} / \Gamma_{t}$ from our measurements. During $t_{w}$, the electrons can relax from the excited to the ground state with a rate $W$. The rate equations describing the model are $\dot{P}_{i}= - \Gamma_{t} P_{i}$ and $\dot{P}_{e} = \Gamma_{e} P_{i} - W P_{e}$. Solving these equations, we find that the probabilities for being in the three states after time $t_{w}$ are given by
\begin{equation}       
 P_{i}(t_{w})= \epsilon_{i} e^{-\Gamma_{t} t_{w}}
\label{eq:PionEqn}
\end{equation}
\begin{equation}       
 P_{e}(t_{w})= \epsilon_{i} \frac{\Gamma_{e}}{\Gamma_{t}} \frac{\Gamma_{t}}{\Gamma_{t} - W}  
 (e^{-W t_{w}}-e^{-\Gamma_{t} t_{w}})
\label{eq:PeEqn}
\end{equation}
and $P_{g}= 1-P_{e}-P_{i}$. It is important to note that in Eq. \ref{eq:PeEqn} the $t_{w}$ dependence of $P_{e}$ depends only on $W$ and $\Gamma_{t}$. In particular, Eq. \ref{eq:PeEqn} has a maximum at $t_{w}= \ln(\Gamma_{t}/W) / (\Gamma_{t} - W)$. We can measure $\Gamma_{t}$ from the $t_{w}$ dependence of $P_{i}$ and then use the $t_{w}$ dependence of $P_{e}$ to determine $W$.

	The third step in the pulse sequence is the real-time readout, shown in \subfig{pulseseq}{d}. We follow Elzerman \etal~\cite{Elzerman2004:SingleShotReadOut} and position the levels so that the excited state is above the Fermi energy of the lead and the ground state is below the Fermi energy. In this configuration, an electron in the excited spin state can quickly tunnel off the dot with rate $\Gamma_{off}$, while the tunneling rate of an electron in the ground state is exponentially suppressed. 
	
	\startsubfig{pulseseq}{f-g} show examples of two types of data. In \subfig{pulseseq}{f} we see that an electron tunnels off during the ionization pulse and back on during the charging pulse. When we enter the readout stage, an electron tunnels off the dot, presumably from the excited state, at a time $t_{off}$ after the end of the charging pulse. Shortly after this, an electron tunnels back onto the empty dot. We call this behavior a `tunnel-off' event. In contrast, in \subfig{pulseseq}{g} we see an electron tunnel off during the ionization pulse, but no electron tunnels on during the charging pulse. Thus the dot is empty entering the readout stage and the first event in this stage is an electron tunneling onto the empty dot. We call this an `ionization event', and measure the time $t_{on}$ between the end of the charging pulse and the time when an electron tunnels onto the dot. The times $t_{off}$ and $t_{on}$ are measured using a triggering and acquisition system described in Ref. \cite{Amasha2006:TunnelingSummary}.

\begin{figure}[!]

\begin{center}
\includegraphics[width=8.0cm, keepaspectratio=true]{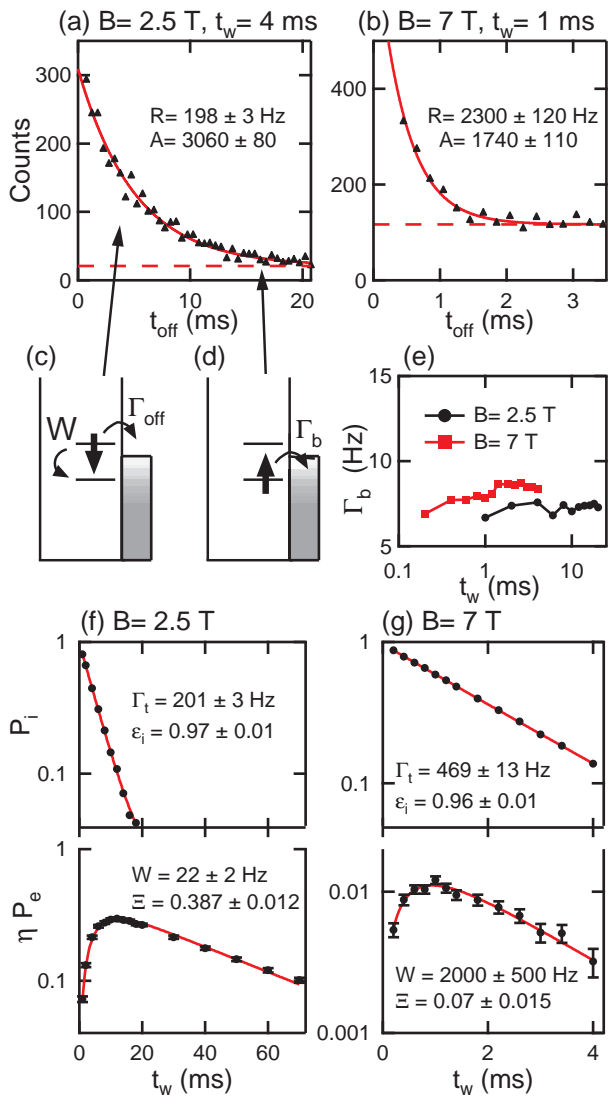}
\end{center}

	\caption{(a)-(b) histograms of $t_{off}$ for tunnel-off events for two different sets of $B$ and $t_{w}$. The exponential is caused by fast tunneling of electrons out of the excited state illustrated in (c) while the offset is caused by slow tunneling out of the ground state as depicted in (d). The rate $R$ is obtained by averaging over the exponentials for several values of $t_{w}$ at each field. This averaging is possible because the readout configuration is the same for all $t_{w}$ at a given field, so $R$ is independent of $t_{w}$. (e) shows the rates $\Gamma_{b}$ extracted from the offsets measured from data like those in (a) and (b). $\Gamma_{b}$ is approximately constant for different $t_{w}$ and between data sets because of the active feedback control discussed in the text. (f)-(g) show measured values of $P_{i}$ and $\eta P_{e}$ at different $B$. The solid lines are fits to Eqs. \ref{eq:PionEqn} and \ref{eq:PeEqn} as discussed in the text.       
}
	\label{fig:PiandPe}
\end{figure}

	For a given $t_{w}$ and $B$ we repeat the pulse sequence $N_{pulse}$ times, where $N_{pulse}$ is typically between $1\times 10^4$ and $1.5\times 10^5$. We histogram the measurements of $t_{off}$ from tunnel-off events; the results are shown in \subfig{PiandPe}{a} and (b) for two different sets of $t_{w}$ and $B$. The data are fit well by an exponential on top of a constant offset. The exponential portion of the data is caused by fast tunneling out of the excited spin state (\subfig{PiandPe}{c}), while the offset is caused by slow tunneling out of the ground spin state (\subfig{PiandPe}{d}). Although the energy of the ground spin state is below the Fermi energy of the lead, there is still a slow rate for tunneling out of the ground state given by $\Gamma_{b}= \Gamma_{off} (1-f(E_{dot}))/(1-f(E_{dot}+\Delta))$, where $f$ is the Fermi function and $E_{dot}<0$ is the depth of the ground state below the Fermi energy of the lead. Our electron temperature allows us to maintain $E_{dot}$ so that $\Gamma_{b} \approx 8$ Hz $<< \Gamma_{off}$. The slow tunneling out of the ground state gives an exponential distribution; however, $1/\Gamma_{b}\approx 100$ ms is so long it appears as a constant offset. 
	
	The ground state tunneling rate $\Gamma_{b}$ is useful for maintaining the stability of the levels in the quantum dot. As noted in previous work \cite{Hanson2005:SpinDepTunnelRates} the energy levels in lithographically defined quantum dots shift over time because of background charge fluctuations in the heterostructure \cite{Pioro-Ladriere2005:SwitchingNoise}. In general, we observe two types of shifts: a slow drift of the levels over time and sudden, large shifts in the position of the energy levels. Our readout method is sensitive to these shifts because it requires that the ground spin state remain below the Fermi energy of the lead while the excited state remains above it. This is a difficult condition to maintain, especially at low $B$ where $\Delta$ is small. Our heterostructure is relatively stable in this regard. In addition, we use active feedback to control the levels in the dot and compensate for these shifts. As noted above, the ground state tunneling rate $\Gamma_{b}$ is a function of $E_{dot}$. We monitor $\Gamma_{b}$ approximately every $15$ s and then adjust a gate voltage to maintain $E_{dot}$ such that $5 \lesssim \Gamma_{b} \lesssim 15 Hz$. We extract $\Gamma_{b}$ from the offset in the histograms in \subfig{PiandPe}{a} and (b). These data are shown in \subfig{PiandPe}{e}. The rate is fairly constant over the range of $t_{w}$ and between data sets, demonstrating the efficacy of our feedback control.  
	
	To measure $P_{e}$ for a given $t_{w}$ and $B$, we need to count the number of times $N_{e}$ that the electron is in the excited state after the charging pulse. For $W \ll \Gamma_{off}$ the exponential decay should have rate $R= \Gamma_{off}$ and the area $A$ under the exponential and above the offset gives $N_{e}$. If $W \approx \Gamma_{off}$, the interpretation is more complicated because an electron in the excited state might relax before it has a chance to tunnel off the dot \cite{Hanson2005:SpinDepTunnelRates}. In this case, the rate of the exponential is the rate at which electrons leave the excited state, namely $R= \Gamma_{off}+W$. Moreover, we only observe the fraction $\eta$ of the electrons in the excited state that tunnel off before they relax given by $\eta= \Gamma_{off}/(\Gamma_{off}+W)$. Thus the area under the exponential and above the offset gives $\eta N_{e}$ and $\eta P_{e}= \eta N_{e}/ N_{pulses}$. Measurements of $\eta P_{e}$ as a function of $t_{w}$ at two different magnetic fields are shown in the lower panels of \subfig{PiandPe}{f} and (g). It is important to note that the information about the relaxation rate $W$ comes from the $t_{w}$ dependence of $P_{e}$, hence the multiplicative factor $\eta$ does not affect our ability to extract $W$.
		
	We can also measure $P_{i}$ as a function of $t_{w}$. We count the number of ionization events $N_{i}$ like those in \subfig{pulseseq}{g} by histogramming the values of $t_{on}$. The distribution is an exponential with no offset and the area underneath the exponential gives $N_{i}$; dividing by $N_{pulses}$ gives $P_{i}$. The upper panels of \subfig{PiandPe}{f} and (g) show examples of $P_{i}$ as a function of $t_{w}$ at two different magnetic fields. We fit $P_{i}(t_{w})$ to Eq. \ref{eq:PionEqn} (solid line) to obtain $\Gamma_{t}$ and $\epsilon_{i}$. While $\Gamma_{t}$ may have some dependence on $B$, it has a much stronger dependence on the tunnel rate set by the gate voltages which may be different for measurements at different $B$. Using the value of $\Gamma_{t}$, we fit $\eta P_{e}(t_{w})$ to Eq. \ref{eq:PeEqn} to find $W$ and the prefactor $\Xi= \eta \epsilon_{i} \Gamma_{e}/\Gamma_{t}$. These fits, shown as the solid lines in the lower panels in \subfig{PiandPe}{f} and (g), give excellent agreement with our data. 
	
	From the upper and lower panels of \subfig{PiandPe}{f} and (g), one can explicitly see the relationship between $P_{i}(t_{w})$ and $P_{e}(t_{w})$ in two different regimes. When $\Gamma_{t}>W$ (\subfig{PiandPe}{f}), $P_{e}$ increases on the time scale of $\Gamma_{t}^{-1}$ and decays on the time scale of $W^{-1}$. When $W>\Gamma_{t}$ (\subfig{PiandPe}{g}), $P_{e}$ increases on the time scale of $W^{-1}$ and decays on the scale of $\Gamma_{t}^{-1}$.  Measuring $\Gamma_{t}$ directly from $P_{i}$ allows us to determine $W$ over a large dynamic range. 

\begin{figure}[!]

\begin{center}
\includegraphics[width=8.0cm, keepaspectratio=true]{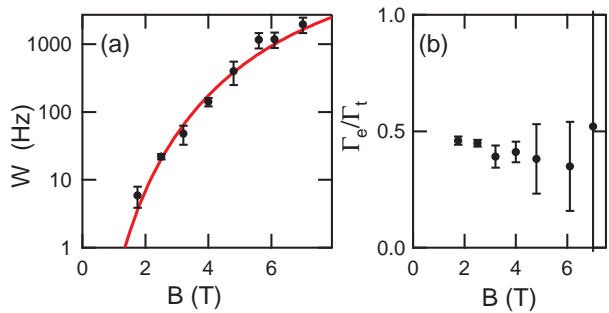}
\end{center}

	\caption{(a) relaxation rate $W$ as a function of magnetic field. The solid line is a theoretical prediction from the work of Golovach \etal~\cite{Golovach2004:PhononInducedDecay} and is discussed in the text. (b) $\Gamma_{e}/\Gamma_{t}$ as a function of magnetic field. From these data we see that $\Gamma_{e}/\Gamma_{t}$ is independent of field and is close to 0.5. The data point at $B= 5.6$ is $\Gamma_{e}/\Gamma_{t}= 2.2 \pm 5.8$ and is not shown. The large errors result from a large error in $\eta$ at this field.         
}
	\label{fig:FitToData}
\end{figure}	
	
	   Using these methods, we measure $W$ as a function of magnetic field. The data are plotted in \subfig{FitToData}{a}. At low fields, the relaxation rate becomes very slow: we measure $\Tone= 170$ ms at $B= 1.75$ T. Golovach \etal~\cite{Golovach2004:PhononInducedDecay} have calculated the relaxation rate caused by spin-orbit coupling between the spin of the electron in the dot and phonons. As inputs to this calculation we use $|g|= 0.38$, which we measure using cotunneling spectroscopy \cite{Kogan2004:ksplitting}, as well as parameters of phonons in GaAs. Also required is $\hbar \omega_{0}$, the energy level spacing of a parabolic potential well that approximates the confining potential of the dot. We estimate this quantity from the energy of the first excited orbital state and find $\hbar \omega_{0}= 2$ meV from transport measurements \cite{MacLean2006:WKB}. The solid line in \subfig{FitToData}{a} shows the results of the calculation using these parameters. We find that a spin-orbit length $\lambda_{SO} = 3~\mu$m gives a curve that agrees well with our data. The contribution of the Dresselhaus and Rashba terms to $\lambda_{SO}$ depends on the orientation of the GaAs crystal with respect to the magnetic field. For our orientation \cite{footnote:orientation} and assuming a symmetric parabolic well, we have $\lambda_{SO}^{-1}= |\lambda_{\alpha}^{-1} -\lambda_{\beta}^{-1}|$ where $\lambda_{\alpha}= \hbar/m^{\ast}\alpha$, $\lambda_{\beta}= \hbar/m^{\ast}\beta$, and $\alpha$ and $\beta$ are the Rashba and Dresselhaus spin-orbit terms respectively in the Hamiltonian $H_{SO} = \beta (-p_x \sigma_x + p_y \sigma_y)+ \alpha (p_x \sigma_y - p_y \sigma_x)$ \cite{Golovach2004:PhononInducedDecay}. This value of $\lambda_{SO}$ is in good agreement with measurements of spin-orbit length scales obtained from antilocalization measurements in quantum dots \cite{Zumbuhl2002:SOCoupling}.
	   
	   We can also extract the value of $\Gamma_{e}/\Gamma_{t}$ as a function of field. From our fits to data such as those in \subfig{PiandPe}{f} and (g), we are able to extract $\eta \epsilon_{i} \Gamma_{e}/\Gamma_{t}$ and $\epsilon_{i}$. We can obtain $\eta$ by noting that $\eta= (R-W)/R $, where $R= \Gamma_{off}+W$ and is measured directly from histograms such as those in \subfig{PiandPe}{a} and (b). This allows us to obtain  $\Gamma_{e}/\Gamma_{t}$ at each value of magnetic field. These values are plotted in \subfig{FitToData}{b}. We see the values are independent of field and very close to $0.5$ as we expect.
	
	 Using our real-time readout technique, we measure $W$ as a function of $B$ down to very low magnetic fields. We find the relaxation rate increases with field, as predicted by theory \cite{Khaetskii2001:ZeemanSF, Golovach2004:PhononInducedDecay}. A quantitative comparison between our measurements and theory gives good agreement, demonstrating that spin-orbit coupling can account for spin relaxation in single quantum dots with one electron. 
	
			We are grateful to V. N. Golovach and D. Loss for discussions and to V. N. Golovach for providing his Mathematica code to perform the calculations. We are also grateful to I. J. Gelfand and T. Mentzel for experimental help. This work was supported by the US Army Research Office under Contract W911NF-05-1-0062, by the National Science Foundation under Grant No.~DMR-0353209, and in part by the NSEC Program of the National Science Foundation under Award Number PHY-0117795.


\begin{thebibliography}{23}
\bibitem[{foo()}]{footnote:orientation}
\bibinfo{note}{The 2DEG is in the $[001]$ plane and the magnetic field is along
  the axis defined by the $[\bar{1}10]$ and $[1\bar{1}0]$ directions}.
\end{thebibliography}

\end{document}